\documentclass[sigconf,anonymous=false]{acmart}
\renewcommand\footnotetextcopyrightpermission[1]{} 
\makeatletter
\renewcommand\@formatdoi[1]{\ignorespaces}
\makeatother

 \usepackage{multirow}
\AtBeginDocument{%
  \providecommand\BibTeX{{%
    \normalfont B\kern-0.5em{\scshape i\kern-0.25em b}\kern-0.8em\TeX}}}

\setcopyright{acmcopyright}
\copyrightyear{2018}
\acmYear{2018}
\acmDOI{10.1145/1122445.1122456}

\begin{document}

\title{Comparing Rationality Between Large Language Models and Humans: Insights and Open Questions}

\author{Dana Alsagheer  }
\affiliation{%
  \institution{University of Houston }
  \city{Houston }
  \state{Texas}
  \country{USA}
  }
\email{dralsagh@cougarnet.uh.edu}

  \author{Rabimba Karanjai }
\affiliation{%
  \institution{University of Houston }
  \city{Houston }
  \state{Texas}
  \country{USA}
  }

  \author{Weidong Shi}
\affiliation{%
  \institution{University of Houston }
  \city{Houston }
  \state{Texas}
  \country{USA}
  }

  \author{Nour Diallo}
\affiliation{%
  \institution{University of Houston }
  \city{Houston }
  \state{Texas}
  \country{USA}
  }
  \author{Yang Lu}
\affiliation{%
  \institution{University of Houston }
  \city{Houston }
  \state{Texas}
  \country{USA}
  }
  \author{ Suha Beydoun}
\affiliation{%
  \city{Houston }
  \state{Texas}
  \country{USA}
  }

   \author{ Qiaoning Zhang}
\affiliation{%
  \country{USA}
  }

\renewcommand{\shortauthors}{Trovato and Tobin, et al.}

\begin{abstract}
This paper delves into the dynamic landscape of artificial intelligence, specifically focusing on the burgeoning prominence of large language models (LLMs). We underscore the pivotal role of Reinforcement Learning from Human Feedback (RLHF) in augmenting LLMs' rationality and decision-making prowess. By meticulously examining the intricate relationship between human interaction and LLM behavior, we explore questions surrounding rationality and performance disparities between humans and LLMs, with particular attention to the Chat Generative Pre-trained Transformer. Our research employs comprehensive comparative analysis and delves into the inherent challenges of irrationality in LLMs, offering valuable insights and actionable strategies for enhancing their rationality. These findings hold significant implications for the widespread adoption of LLMs across diverse domains and applications, underscoring their potential to catalyze advancements in artificial intelligence.
\end{abstract}.



\maketitle
\thispagestyle{empty}
\section{Introduction}


Large language models (LLMs) have emerged as a transformative advancement in artificial intelligence, demonstrating remarkable proficiency in text manipulation, encompassing tasks from question answering to nuanced reasoning~\cite{hadi2023large,naveed2023comprehensive}. Their extensive training on massive text datasets equips them with a vast knowledge repository from factual information to abstract principles governing the physical world. This breadth of knowledge empowers LLMs to perform sophisticated language tasks with unprecedented finesse.

The increasing sophistication of LLMs highlights the pressing need to delve deeper into their learning mechanisms and scrutinize the intricacies of their challenges. LLMs evoke admiration and scrutiny, with proponents citing their potential for general intelligence due to their extensive training on vast datasets. However, skeptics raise concerns about their limitations in fully grasping human-like language and semantics. This ongoing discourse underscores the importance of rigorous evaluation methods to assess these models' capabilities accurately.

Human rationality epitomizes intelligent behavior, characterized by analytical thinking and decision-making aligned with normative standards~\cite{chater2001human}. Evaluating the rationality of LLMs involves examining their decision-making processes and problem-solving abilities across diverse domains. Rationality, a multifaceted concept influenced by context, encompasses epistemic rationality based on evidence and instrumental rationality serving personal objectives. As we strive to deepen our understanding of LLM rationality, constructing comprehensive evaluation frameworks becomes essential to capture the complexity of their decision-making mechanisms.

Reinforcement Learning from Human Feedback (RLHF) stands at the forefront of advancements in training and refining LLMs, enhancing their ability to interpret and execute human instructions effectively~\cite{ouyang2022training,wang2023aligning}. Through RLHF, LLMs can discern user intentions and refine their responses based on past interactions, transcending their traditional role as mere auto-completion tools~\cite{lambert2023history}.

Utilizing machine learning models built upon human preferences, particularly those employing RLHF for optimization can significantly impact user interactions with resulting systems~\cite{krugel2023chatgpt,lambert2023history}. Concerns persist about the stability and robustness of RLHF-trained LLMs, prompting investigations into potential effects on users' moral judgments, rational thought processes, and susceptibility to biases.

Our study marks a pioneering effort to analyze the irrationality within LLMs and juxtapose it against human rationality. Our contributions are twofold:
\begin{itemize}
  \item Comparative analysis of rationality performance between humans and LLMs: The study provides insights into how well LLMs align with human rationality across diverse contexts and decision-making scenarios through rigorous experimentation and evaluation.

  \item Addressing irrationality and proposing solutions to enhance transparency and auditing: The paper delves into the challenge of irrationality in LLMs caused by human feedback and offers solutions to bolster transparency and auditing, aiming to pave the way for developing more rational models.
\end{itemize}
 The paper's organization is as follows: Section 2 reviews related work, Section 3 presents the method, Section 4 discussion, and Section 5 concludes the paper.


\begin{table*}[]
\centering
\caption{Example of the Rationality Test.}

\begin{tabular}{|c|l|}
\hline
\textbf{Type}                         & \multicolumn{1}{c|}{\textbf{Example}}                                                                                                                                                                                                                                                                                                                                                                                                                                                                                     \\ \hline
\textbf{Wason Selection Task}         & \begin{tabular}[c]{@{}l@{}}Instructions: In this task you will be shown four cards with a rule beneath them.\\  Each card has two sides, but you will only see one.\\ Your job is to click the two cards you need to turn over to decide\\ whether the rule is true or false.\\ Scenario: Suppose each card below has a letter on one side and a number on the \\ other.\\ Rule: If a card has a V on one side, it has an even number on the other.\\ Face Up Cards: V; S; 2; 5\end{tabular}     \\ \hline
\textbf{Conjunction Fallacy Task}     & \begin{tabular}[c]{@{}l@{}}Scenario: Suppose each card below has a decision on one side and a height on\\ the other.\\ Rule: You must be at least 5 feet tall to ride a roller coaster.\\ Face Up Cards: Can Ride Roller coaster; Cannot Ride Roller coaster; 5 ft Tall; 4 ft Tall\end{tabular}                                                                                                                                                                                                                    \\ \hline
\textbf{Stereotype Base Rate Neglect} & \begin{tabular}[c]{@{}l@{}}Scenario: In a study 1000 people were tested. Among the participants were\\ three who lived in a condo and 997 who lived in a farmhouse. Kurt is a randomly\\ chosen participant in this study.\\ Description: Kurt works on Wall Street and is single. He works long hours, and\\ wears Armani suits to work. He likes wearing shades.\\ What is more likely?\\ Option 1: Kurt lives in a condo\\ Option 2: Kurt lives in a farmhouse\\\end{tabular} \\ \hline
\end{tabular}
   \label{tab:my_label}
\end{table*}

\section{Related work }
\subsection{Reinforcement Learning from Human Feedback}
Reinforcement learning from human feedback (RLHF) is a robust method to enhance LLMs by harmonizing them with human objectives. Despite its widespread use, a notable need exists for more transparency regarding the internal workings and limitations of RLHF. Documentation on RLHF reward models, pivotal for achieving superior results, remains sparse. This gap underscores the necessity for further research and transparency concerning RLHF reward models~\cite{lambert2023history,thrun2000reinforcement}. 
In a related study, in~\cite {casper2023open,hendrix2023intricacies}, flaws in RLHF's approach to training AI systems to align with human goals are examined. This study identifies open problems, proposes techniques for improvement, and advocates for auditing standards to bolster oversight of RLHF systems. These endeavors underscore the importance of adopting a comprehensive approach to developing safer AI systems, emphasizing the imperative for thoroughly examining and enhancing RLHF methodologies.
\subsection{Cognitive and Reasoning in Artificial intelligence }
Recent research efforts, exemplified by studies such as \cite{binz2023using}, have shed light on the strengths and weaknesses of Language Models (LLMs), employing insights from cognitive psychology to delve into their operational mechanisms. While LLMs like ChatGPT exhibit remarkable proficiency across diverse tasks, they also unveil vulnerabilities, particularly in domains requiring causal reasoning. In parallel, another notable work \cite{9793310} has examined the interaction between AI and human cognition. This study investigates how AI can be effectively harnessed to enhance reasoning abilities and address mental health challenges. Furthermore, it seeks to identify specific problems within the domain of mental health that can be addressed through AI reasoning methodologies. Such research endeavors aim to advance our understanding of cognitive processes and facilitate innovative solutions to promote mental well-being.
\subsection{Rationality}
Rationality encompasses the broader aspect of decision-making and behavior, while reasoning focuses on the mental processes involved in concluding \cite{manktelow2012thinking}. This cognitive ability is indispensable across a spectrum of intellectual pursuits, encompassing problem-solving, decision-making, and critical thinking \cite{chater2001human,russell2016rationality}. Seminal works in psychology, such as those by Wason \cite{wason1972psychology} and Wood \cite{ad-wood-2003}, underscore the pivotal role of reasoning in comprehending human cognition and behavior. Rationality epitomizes intelligent conduct characterized by analytical thinking and the ability to make decisions that either maximize expected utility or adhere to probabilistic principles, thus aligning with normative decision-making standards. The significance of rationality spans diverse scenarios, ranging from mundane choices like grocery shopping to consequential decisions like retirement planning. Empirical evidence indicates that varying levels of rationality correlate with real-world outcomes; diminished decision-making competence has been associated with issues such as juvenile delinquency in adolescents \cite{gigerenzer1996reasoning}.
Moreover, rationality is a multifaceted concept influenced by contextual factors. Optimal decisions may vary based on individual or group interests, giving rise to the notion of relative rationality \cite{o2016rationality}. Additionally, rationality extends beyond traditional decision-making domains, influencing areas such as religious beliefs and susceptibility to misinformation.

Despite previous efforts, there remains a gap in research regarding examining irrationality and its impact on refining models through human feedback. This unexplored aspect highlights the need for further investigation to understand how irrational human feedback might affect the effectiveness and reliability of models enhanced with RLHF. Moreover, there needs to be more research quantifying rationality in LLMs and understanding their decision-making processes and limitations, which is essential for bolstering the robustness and applicability of AI models in real-world scenarios.

\section{METHOD}

\subsection{Assessing Irrationality}
Our investigation aimed to assess LLMs rationality through specific tasks and compare it with human rationality using rationality tests. These tests included:
\begin{itemize}
\item Wason Selection Task: This task involves applying conditional logic rules by selecting cards to test the validity of a given rule, often revealing confirmation bias. It comprises eleven questions. The Wason selection task provides a window into the complex interplay between logical reasoning, cognitive biases, and decision-making processes. Researchers can deepen their understanding of participants' rationality and mental functioning by studying participants' performance on this task under different conditions and interventions. 
\item Conjunction Fallacy Test: It evaluates individuals' tendency to overestimate the likelihood of two events occurring together despite each event having a lower probability individually. Participants are presented with seven questions where they encounter two statements, one of which is a conjunction that seems more believable but is less probable. This phenomenon highlights the impact of cognitive biases on rational decision-making. Understanding this fallacy provides valuable awareness of the constraints of human reasoning, especially in situations characterized by uncertainty~\cite{tversky1988rational}.

\item Stereotype Base Rate Neglect: Participants analyze scenarios containing base rate information and stereotypes to determine the group a described person belongs to. Conflict trials highlight their tendency to overlook base rate information when it conflicts with stereotypes. This section comprises eleven questions.
\end{itemize}
Each task was selected to unveil underlying rationality factors and offer profound insights into LLMs' cognitive capabilities.
 \tablename~\ref{tab:my_label} illustrates the Examples of questions used in each section. 
By applying these tasks to assess ChatGPT's responses, we aim to gain insights into its rationality and decision-making processes across various scenarios, providing valuable insights into its capabilities and limitations.




\subsection{Humans Data }

The methodology employed to measure irrationality in this study was thorough and systematic. In our study, we utilized data from a method akin to a previous project, where they recruited 300 participants from the Georgia Institute of Technology and the Atlanta community. All participants met specific criteria, including providing informed consent, being native English speakers who learned English before age 5, and falling within the age range of 18 to 35. Values deviating more than 3.5 standard deviations from the mean were treated as missing, except for the Wason selection task, where five positive outliers were retained due to a pronounced floor effect in the scores. The study aimed to elucidate the relationship between broad cognitive abilities and rationality using latent variable analyses, which offer a more comprehensive perspective than individual tests by assessing fluid intelligence, working memory capacity, and attention control alongside measures of rationality, including tasks such as the Wason selection task, base rate neglect, and conjunction fallacy tests~\cite{burgoyne2023understanding}.

\subsection{Online Humans Data }

Our research methodology extended into the online realm to ensure consistency and comparability between human participants and ChatGPT. We distributed 50 online questionnaires targeting individuals with advanced educational backgrounds, specifically those holding master's degrees. Our selection process was meticulous, aiming to align with the academic profile of participants recruited from the Georgia Institute of Technology and the Atlanta community. Unlike Georgia's criteria, which focus solely on language and age considerations, we sought participants with advanced educational qualifications. This expansion enabled us to explore how individuals with similar academic backgrounds approached the Wason selection task remotely, offering insights beyond traditional face-to-face interactions. The online questionnaire mirrored tasks administered to the primary participant group, ensuring consistency in task administration across both online and in-person settings.
\subsection{ChatGPT Data }
Our methodology for assessing rationality begins by leveraging two distinct language models: ChatGPT and Gemini. We utilize questions from a standardized test similar to those used in evaluations at the Georgia Institute of Technology. Initially, we observe consistent rationality outcomes across both models. Seeking to refine our approach, we conduct 350 unique API calls exclusively to the ChatGPT platform using the same standardized questions, aiming to elicit varied outcomes.
\subsection{Result  }
In this section, we present the experimental findings and compare the performance of ChatGPT with that of the human participants.

\begin{figure}[h]
  \centering
  \includegraphics[width=\linewidth]{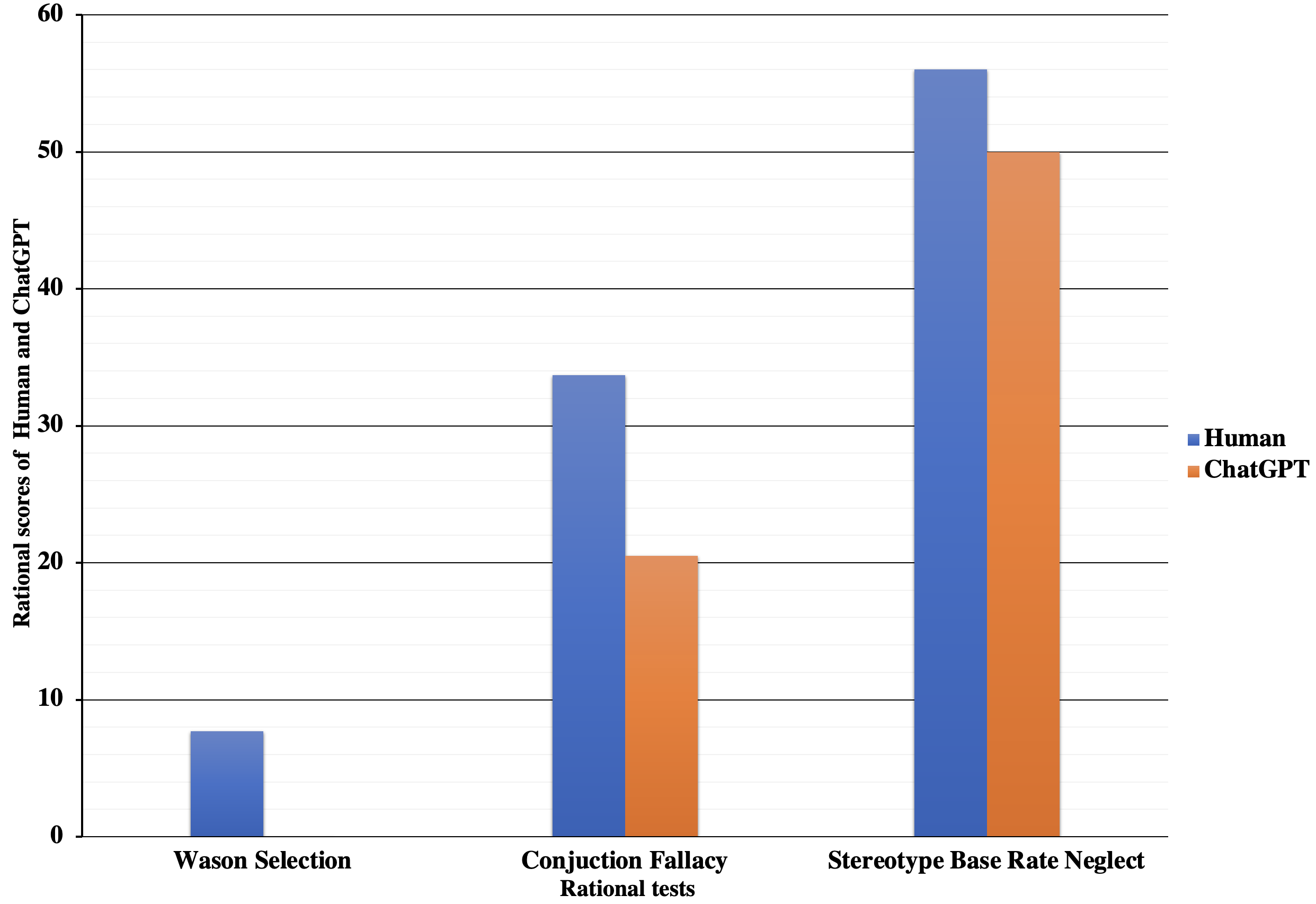}
  \caption{Illustrates the compression between humans and ChatGPT.}
 \label{fig:CHA}
\end{figure}

\begin{enumerate}
    \item\textbf {Wason selection task:} The evaluation results revealed a stark contrast between ChatGPT's performance and that of the human participants. ChatGPT consistently provided incorrect answers, indicating a significant deficiency in her ability to comprehend and apply the logic or reasoning required for the task. Moreover, her accompanying explanations revealed a fundamental misunderstanding of the task's objectives, further highlighting her inability to grasp the essential principles involved. Consequently, ChatGPT received a total score of 0.5, reflecting a complete divergence from the correct solutions. In contrast, the human participants, whether online or face-to-face, exhibited a more varied range of performance. While some individuals achieved relatively low scores (ranging from 4\% to 15\%), the majority demonstrated a higher level of understanding than ChatGPT. This was evident in their ability to provide reasoned explanations for their answers, offering valuable insights into their decision-making processes and the underlying rationale guiding their choices. Despite variations in individual performance, the human participants' engagement with the task and their capacity to articulate their reasoning highlighted a level of comprehension that was notably absent in ChatGPT's responses. Overall, while both ChatGPT and the human participants struggled with the task to varying degrees, the ability of the human participants to comprehend and articulate their reasoning suggests a higher level of cognitive engagement and understanding. This underscores the complexity of the task and emphasizes the importance of mental abilities such as rationality and information relevance assessment, which may vary between artificial intelligence systems and human participants.

 \item\textbf{Conjunction fallacy:} Although there was a slight performance improvement, the results remain relatively low. Human participants and LLMs showed subpar performance, with LLMs scoring 28\% and humans scoring 33.7\%. Online human participants achieved a comparatively better score of 46\%. These findings underscore the widespread presence of this cognitive bias across various cognitive systems, emphasizing its persistent challenge.

\item\textbf{Base rate neglect:} Indeed, while human participants scored 56\%, online humans scored 60\%, and ChatGPT scored 50\% on the test assessing rationality. Notably, ChatGPT achieved a score remarkably close to that of humans. This proximity in performance can be attributed to several factors inherent to ChatGPT functioning. Firstly, LLMs can process vast prior knowledge, drawing upon extensive datasets and linguistic patterns to inform their predictions and decisions. This broad knowledge base allows LLMs to consider a wide array of statistical trends and historical data when confronted with new information, thereby mitigating the effects of base rate neglect. Secondly, LLMs are not susceptible to cognitive biases like humans are. While humans may prioritize new, event-specific information over broader statistical trends, LLMs are programmed to weigh all available data objectively without succumbing to such biases ~\cite{dasgupta2022language,macmillan2023ir}. Additionally, logical rules and algorithms guide LLMs' decision-making processes, ensuring consistency and accuracy in their assessments. Therefore, despite the task's inherent complexity, LLMs demonstrate a level of rationality comparable to that of humans, as evidenced by their performance on the test.

The analysis of test results in  \figurename~\ref{fig:CHA} reveals significant challenges humans and LLMs face in achieving high rationality scores, emphasizing the critical need to assess the effectiveness of human feedback. Despite concerted efforts to provide rational feedback, human participants often display irrational tendencies, potentially introducing biases that may skew the evaluation process. Furthermore, LLMs heavily rely on existing knowledge, which could hinder their ability to adapt to novel scenarios and incorporate human feedback efficiently. Overcoming this challenge requires ensuring that humans can offer rational feedback, thus mitigating the perpetuation of irrational models.

The strategic integration of an online platform alongside careful participant selection has proven to be a catalyst for improving the test results. This approach has played a pivotal role in enhancing task performance, exemplified by significant advancements observed, particularly in tasks like the Wason selection task. 
Such results underscore the intrinsic value of the online format in refining methodological approaches. Through the seamless integration of an online questionnaire, we have expanded the breadth of perspectives captured, enhancing the robustness of our study outcomes. This holistic strategy underscores the indispensable role of the online platform in our research methodology, emphasizing its crucial inclusion for future assessments to ensure thoroughness and validity.
\end{enumerate}

\section{disccision }


 \subsection{Open Quistions }

\textbf{What are some potential implications of incorporating human biases and heuristics into artificial intelligence systems, particularly in light of acknowledging human irrationality?}

 Incorporating human biases and heuristics into artificial intelligence (AI) systems offers benefits and challenges, especially when considering human irrationality. While efforts have been made to develop rational AI agents, it's crucial to recognize that humans often deviate from rational behavior due to cognitive biases. Integrating these biases into AI systems creates a more accurate representation of human decision-making, enabling more effective human-machine interaction. Understanding and integrating human irrationality can lead to AI systems better adapting to real-world scenarios and improving their performance across various tasks. Moreover, studying human irrationality within AI research contributes to a deeper understanding of human behavior and cognition, fostering advancements in psychology and artificial intelligence. However, this recognition of human irrationality underscores the complexity of designing AI systems that can emulate human-like decision-making processes while highlighting the potential benefits of leveraging human biases in AI design and development\cite{gigerenzer1996reasoning,kahneman1982judgment}.

\textbf{How can we establish criteria for selecting individuals to provide feedback to the Large-Language Model (LLM) in reinforcement learning settings?}

 Selecting feedback providers for Large Language Models (LLMs) in reinforcement learning settings requires careful consideration of human variability in performance. Tests like the Wason selection task, where only a fraction of participants answer correctly, highlight this variability. Choosing feedback providers based on demonstrated rationality in relevant tasks is crucial to ensure the model receives rational and practical guidance. Exploring strategies for identifying feedback providers with a consistent track record of rational decision-making can significantly improve the quality of feedback and enhance the LLM’s learning process. Additionally, intentional integration of human biases and heuristics into AI systems can prove advantageous in specific scenarios, provided they are carefully managed and aligned with the model’s objectives. However, differentiating between unintentional biases due to training data and intentional integration of biases is essential to mitigate adverse outcomes and enhance decision-making\cite{macmillan2023ir,wen2019modelling}.

\textbf{How can auditing and transparency mechanisms be effectively implemented to ensure rationality in developing and deploying AI systems, particularly within human feedback-driven training methodologies?}

 Human feedback plays a critical role in constructing rational models in refining training methodologies for AI systems. Implementing auditing and transparency mechanisms is essential to ensure accountability and mitigate risks within the Large Language Model and reinforcement learning with human feedback (RLHF) governance framework. Transparency is pivotal in enhancing and evaluating human feedback, especially as societal scrutiny intensifies around responsible governance frameworks for AI systems. Ensuring transparency and accountability in the feedback process is paramount. Several key elements should be disclosed, including the description of the pretraining process, selection, and training of human evaluators, process for selecting feedback examples, types of human feedback used, and quality assurance measures. By disclosing these aspects of human feedback, transparency is bolstered, enabling stakeholders to understand better the feedback process's impact on the model’s development and performance. However, challenges remain in incorporating such standards into AI governance norms and regulations, requiring concerted efforts from stakeholders across various domains\cite{casper2023open}.

\textbf{Can the large-language model-based RLHB approach be built using Universal democratic norms?}

 Developing a large-language model-based RLFB approach rooted in universal democratic norms presents a profound challenge. Consider an AI conversational model tailored for educational contexts, trained using Reinforcement Learning from Human Feedback (RLHF). Users may express preferences, like avoiding encounters with language they perceive as discriminatory. Sen’s theorem underscores the complexity of crafting an RLHF model through democratic means that honor each individual’s private inclinations. This dilemma exposes the formidable endeavor of attaining alignment with the diverse intentions of all users in Artificial General Intelligence. While it’s conceivable to fine-tune AI to match individual user preferences closely, achieving comprehensive alignment across diverse user cohorts or tasks remains inherently constrained. Thus, while AI can be customized to accommodate particular preferences or domains, achieving universal alignment in AI LLMs through RLHF methodology remains challenging and may require transparent communication and meticulous attention during the reinforcement learning process \cite {mishra2023ai}.

\subsection{Limitations}
Developing a large-language model-based RLFB with universal democratic norms presents a significant challenge due to the limitations of rational decision-making. Users often express preferences beyond rational considerations, complicating alignment with diverse intentions, particularly in artificial general intelligence (AGI). Designing RLFB models that respect individual inclinations within the rationality framework presents a formidable challenge in achieving universal alignment across AI Large Language Models (LLMs) \cite{mishra2023ai}. Moreover, acknowledging human irrationality adds another layer of complexity to developing AI systems. Despite endeavors to create rational LLMs, human cognitive biases frequently lead to deviations from rational behavior, necessitating their integration into AI systems. Understanding human irrationality benefits psychology and enhances artificial intelligence, emphasizing the challenge of designing AI systems that emulate human-like decision-making while leveraging human biases \cite{kahneman1982judgment,gigerenzer1996reasoning}.
Conversely, assessing Natural Language Generation (NLG) systems via human ratings frequently needs to acknowledge the creative dimensions of human cognition. While aggregating ratings across annotators aims to encapsulate collective preferences, it often overlooks the subtleties of creativity. Despite continuous efforts to enhance NLG evaluation techniques, these constraints persist. Thus, there is a pressing need for the creation of inventive evaluation methodologies that not only acknowledge but also embrace human irrationality, recognizing its potential to spur creativity, for a more precise assessment of NLG systems \cite{ethayarajh-jurafsky-2022-authenticity}.

\section{Conclusion }

Large Language Models (LLMs) have made remarkable strides in certain aspects of rational thinking and complex cognitive processes. However, this study underscores the necessity of subjecting LLMs to rationality tests, revealing challenges in their ability to answer simple questions accurately. This highlights the need to explore LLMs' natural language processing and narrative generation capabilities further. Understanding the mechanisms underlying LLMs' rationality and refining evaluation methodologies is essential to overcome these challenges. Collaboration across disciplines and the adoption of innovative approaches are paramount to fully unlocking the potential of LLMs in reasoning and advancing artificial intelligence.

Furthermore, rationality and irrationality play pivotal roles in AI, necessitating a multidisciplinary perspective. Recognizing the potential benefits of irrational behavior in specific contexts drives ongoing research into methods for effectively interacting with irrational agents. In human-AI interaction, accommodating human irrationality is crucial, given humans' integral role in providing feedback. While cognitive biases may enhance the performance of artificial agents, system design must be adaptable to human irrationality. Exploring how the rationality of artificial agents influences the dynamics of human-AI interaction underscores the importance of addressing lingering questions as AI becomes increasingly integrated into daily life.

\bibliographystyle{ACM-Reference-Format}
\bibliography{sample-base}








\appendix

\end{document}